\documentclass[aps,pra,twocolumn,superscriptaddress,%
floatfix]{revtex4-1}
\usepackage{amsmath,amssymb,bm,epsf,graphicx,epsf,subfigure}

\newcommand{\avg}{\mathrm{avg}}
\newcommand{\bk}{\mathbf{k}}

\newcommand{\chain}{\mathrm{chain}}
\newcommand{\disc}{\mathrm{disc}}

\newcommand{\half}{{\textstyle\frac{1}{2}}}
\newcommand{\imp}{\mathrm{imp}}
\newcommand{\kept}{\mathrm{kept}}
\renewcommand{\L}{\Lambda}
\newcommand{\pdag}{\phantom{\dag}}

\newcommand{\Sea}{S_e^{\avg}}
\newcommand{\Sei}{S_e^{\imp}}
\newcommand{\Sem}{S_e^{\max}}
\newcommand{\tveps}{\tilde{\veps}}

\newcommand{\veps}{\varepsilon}

\begin{document}

\title{Long-range entanglement near a Kondo-destruction quantum critical point}
\author{Christopher Wagner}
\affiliation{Department of Physics, University of Florida, Gainesville, Florida
32611-8440, USA}
\author{Tathagata Chowdhury}
\affiliation{Department of Physics, University of Florida, Gainesville, Florida
32611-8440, USA}
\affiliation{Institut f\"{u}r Theoretische Physik, Universit\"{a}t zu K\"{o}ln,
Z\"{u}lpicher Strasse 77a, 507937 K\"{o}ln, Germany}
\author{J.\ H.\ Pixley}
\affiliation{Department of Physics and Astronomy, Center for Materials Theory,
Rutgers University, Piscataway, NJ 08854 USA}
\affiliation{Condensed Matter Theory Center and the Joint Quantum Institute,
Department of Physics, University of Maryland, College Park,
Maryland 20742-4111, USA}
\author{Kevin Ingersent}
\affiliation{Department of Physics, University of Florida, Gainesville, Florida
32611-8440, USA}
\date{\today}

\begin{abstract}
The numerical renormalization group is used to study quantum entanglement in
the Kondo impurity model with a pseudogapped density of states
$\rho(\veps)\propto|\veps|^r$ ($r>0$) that vanishes at the Fermi energy
$\veps=0$. The model features a Kondo-destruction quantum critical point (QCP)
separating a partially screened phase (reached for impurity-band exchange
couplings $J>J_c$) from a local-moment phase ($J<J_c$).
The impurity contribution $\Sei$ to the entanglement entropy between a region
of radius $R$ around the magnetic impurity and the rest of the host system
reveals a characteristic length scale $R^*$ that distinguishes a regime
$R\ll R^*$ of maximal critical entanglement from one $R\gg R^*$ of weaker
entanglement. Within each phase, $\Sei$ is a universal function of $R/R^*$
with a power-law decay for $R/R^*\gg 1$. The entanglement length scale $R^*$
diverges on approach to the QCP with a critical exponent that depends only
on $r$.
\end{abstract}


\maketitle

Advances in quantum information have led to the quantification of entanglement
\cite{Janzing:09}, which has helped develop new fundamental concepts in
condensed matter physics \cite{Amico:08}. The entanglement entropy $S_e$
characterizes the entanglement of a pure state of a system with respect
to a partition into two subsystems $A$ and $B$. If $S_e>0$, then a measurement
that collapses the state of $A$ will also collapse the state in $B$, whereas if
$S_e=0$, such a measurement on $A$ will not affect $B$.
The entanglement of a subsystem has recently been measured in ultra-cold atomic gases \cite{Islam:15}, making it
experimentally relevant to ask how the entanglement scales with the length $l$
of the smaller subsystem in $d$ spatial dimensions.
Certain eigenstates can be classified by an ``area law''
$S_e \sim l^{d-1}$ (applicable, e.g., to various ground states
\cite{Eisert:10}) or by a ``volume law'' $S_e \sim l^d$ (typical
for highly excited states in a thermal system \cite{Nandkishore:14}).
The existence of a Fermi surface can impart a logarithmic correction to the
area law for ground states of fermionic systems at finite density, i.e.,
$S_e \sim l^{d-1}\log l$ \cite{Gioev:06}. In more exotic phases that lack a
local order parameter, $S_e = a l - \gamma+\dots$ describes a
long-range-entangled ground state with a universal area-law offset $\gamma$
due to topological order in $d=2$ \cite{Kitaev.06,Levin.06}.

Entanglement entropy has been particularly successful at characterizing the
ground states of quantum impurity models, in which a local dynamical
degree of freedom can be screened via entanglement with a dense set of host
energy levels.
For example, the Kondo effect is an inherently quantum-mechanical phenomenon
due to its singlet ground state \cite{Hewson:93}. It is therefore natural to
expect that the size of the Kondo screening cloud dictates the spatial range
of entanglement, a picture that has been confirmed for an interacting spin
chain described by the same effective low-energy theory as the Kondo model
\cite{Sorensen:07,Affleck:09}. However, a direct observation in
the degrees of freedom of the original model has hitherto been lacking.
Moreover, in situations where the Kondo effect can be driven critical at a
continuous quantum phase transition \cite{Withoff:90,Bulla:97,
Gonzalez-Buxton:98,Ingersent:02,Fritz:04,Glossop:11,Schneider:11,Pixley:12,
Sengupta:00,Zhu:02,Zarand:02,Kircan:04,Glossop:05,Glossop:07,Pixley:15},
the fate of the Kondo screening cloud and the spatial structure of
entanglement are both poorly understood. Is entanglement long ranged at a
Kondo-destruction quantum critical point (QCP), despite the impurity
becoming asymptotically free at low temperatures? This question is relevant
for certain heavy-fermion compounds---such as CeCu$_{6-x}$Au$_x$
\cite{Schroder:00}, YbRh$_2$Si$_2$ \cite{Paschen:16} and CeRhIn$_5$
\cite{Shishido:05}---that are believed to exhibit a Kondo-destruction QCP
concomitant with a jump in the Fermi-surface volume. Entanglement entropy
can provide crucial insights into the nature of the many-body ground state
near such a bulk QCP.

In this Letter, we show that the numerical renormalization group (NRG) can be
used to accurately calculate the entanglement in the ground state of a
spin-$\half$ magnetic impurity in a metallic or semimetallic host. Previously,
we have investigated the ``local'' entanglement between such an impurity and
its host, taking the impurity alone to form subsystem $A$ \cite{Pixley:15}.
Here, we instead compute $\Sei(R)$, the impurity contribution to the
entanglement entropy between a region of radius $R$ about the impurity site
and the rest of the system. For a metal, where the impurity spin becomes fully
screened at temperatures $T\ll T_K$ (the Kondo temperature), we directly\
confirm the previously deduced \cite{Sorensen:07,Affleck:09} scaling of $\Sei$
with $R/R_K$, where $R_K \propto 1/T_K$ is believed to be the characteristic
size of the many-body Kondo screening cloud.

Our main results are for the pseudogap Kondo model \cite{Withoff:90}, which
features a Kondo-destruction QCP at an impurity-band exchange coupling
$J=J_c$ separating a partially screened Kondo phase ($J>J_c$) from a
local-moment phase ($J<J_c$) in which there is no \emph{static\/} Kondo
effect. Each phase reveals a length scale $R^*$ such that for $R\ll R^*$,
$\Sei$ takes its maximal value, a signature of strong entanglement associated
with the QCP. In the Kondo phase, $\Sei$ decreases for $R\gg R^*,$ but (in
contrast to the conventional metallic case) remains nonzero even for
$R\to\infty$ due to the incomplete screening of the impurity
\cite{Gonzalez-Buxton:98}. In the local-moment phase, the strong entanglement
for $R\ll R^*$ evidences a \emph{dynamical\/} Kondo effect, but $\Sei$
drops toward zero for $R\gg R^*$.
In both phases, $\Sei$ obeys universal scaling in terms of $R/R^*$ with a
power-law decay for $R/R^*\gg 1$ described by a non-integer exponent.
On approach to the QCP, the entanglement length diverges like
$R^*\sim |J-J_c|^{-\nu}$, leading to a maximal, scale-invariant
entanglement extending from the impurity throughout the entire system.

\begin{figure}
\centering
\includegraphics[width=0.6\columnwidth]{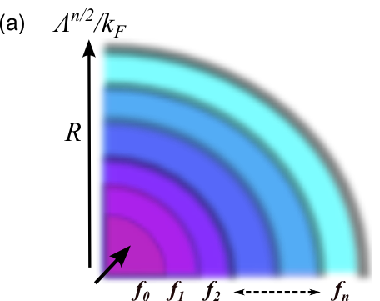} \\
\includegraphics[width=0.7\columnwidth]{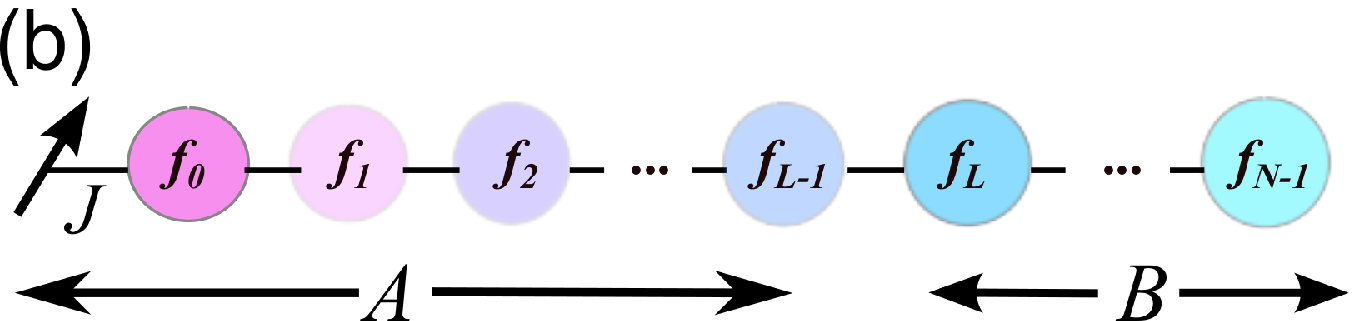}
\caption{\label{fig:schematic}
NRG representation of the Kondo model as a tight-binding Wilson chain of
$N$ sites coupled at one end to an impurity spin.
(a) In real space, Wilson site $n$ corresponds to a spherically symmetric band
state with a radial probabilty density peaked at a radius
$\propto k_F^{-1} \L^{n/2}$ from the impurity.
(b) The entanglement entropy $S_e(J,L,N)$ is found by splitting the mapped
system into subsystems $A$ (the impurity and the first $L$ Wilson
sites) and $B$ (the remaining $N-L$ sites).
}
\end{figure}

\emph{Model.}
We consider the spin-$\half$ Kondo Hamiltonian
\begin{equation}
\label{eq:Ham}
H = \sum_{\bk,\sigma} \veps_{\bk} \,
    c_{\bk\sigma}^{\dag} \, c_{\bk\sigma}^{\pdag}
    +  \frac{J}{2N_{\bk}} \, \mathbf{S}_{\imp} \cdot \!\!\!\!
                \sum_{\bk,\bk',\sigma,\sigma'} \!
    c_{\bk\sigma}^{\dag} \, \bm{\sigma}_{\sigma\sigma'} \:
    c_{\bk'\sigma'}^{\pdag},
\end{equation}
where $c_{\bk\sigma}$ destroys a band electron of energy $\veps_{\bk}$ and
spin $z$ component $\sigma=\pm\half$, $N_{\bk}$ is the number of $\bk$ values
(i.e., the number of host unit cells), $J$ is the local exchange
coupling between band electrons and the impurity spin $\mathbf{S}_{\imp}$,
and $\bm{\sigma}$ is a vector of Pauli matrices. We consider a density of
states of the (highly simplified) form
\begin{equation}
\label{eq:dos}
\rho(\veps)
= N_{\bk}^{-1} \sum_{\bk} \delta(\veps - \veps_{\bk})
= \rho_0 |\veps/D|^r\Theta(D-|\veps|),
\end{equation}
where $D$ is the half-bandwidth and $\Theta(x)$ is the Heaviside function.
The model has a rich phase diagram that crucially depends on the band exponent
$r$ \cite{Gonzalez-Buxton:98}. The case $r=0$ corresponds to the conventional
Kondo problem in a metal \cite{Hewson:93}.
For semimetals with $0<r<\half$, the above-mentioned Kondo-destruction QCP
occurs at $J=J_c>0$. At this interacting QCP, the system exhibits a critical
impurity spin response characterized by nontrivial, $r$-dependent exponents
\cite{Ingersent:02}.

We consider the impurity-induced change in the entanglement entropy,
defined as $\Sei(J,R) \equiv S_e(J,R)-S_e^{(0)}(R)$.
Here, $S_e(J,R)$ is the entanglement entropy of the combined impurity-band
system with subsystem $A$ consisting of the impurity plus that part of the
band within radius $R$ of the impurity site, and $S_e^{(0)}(R)$ is the
entanglement entropy of the band alone when partitioned at the same radius $R$
[see Fig.~\ref{fig:schematic}(a)]. Since the exchange coupling in Eq.\
\eqref{eq:Ham} is spherically symmetric, the impurity affects only the
$s$-wave band degrees of freedom, and for purposes of calculating
impurity-induced properties, the problem reduces to one (radial) dimension.
After this reduction, one has \cite{Gioev:06,Swingle:10,Ding:12}
$S_e^{(0)}(R) \sim \log R$ rather than the full three-dimensional
behavior $S_e^{(0)}(R) \sim R^2 \log R$.

\emph{Computational method.}
We study the radial Kondo model using the NRG \cite{Wilson:75,Bulla:08} as
modified to treat a power-law density of states \cite{Gonzalez-Buxton:98}.
The Hamiltonian is mapped onto a semi-infinite tight-binding ``Wilson chain''
of sites labeled $n = 0$, $1$, $2$, $\ldots$, coupled to the impurity via site
$0$ only. A discretization parameter $\L > 1$ introduces a separation of energy
scales that causes the nearest-neighbor hopping coefficients to decay
exponentially as $t_n \sim D\L^{-n/2}$ and allows iterative diagonalization
of Kondo Hamiltonians $H_M$ having finite Wilson chains of length $M$ with
$M=1$, $2$, $\ldots$, $N$.

To quantify entanglement, the system described by $H_N$ is divided into a
subsystem $A$ comprising the impurity and the first $L$ chain sites
($0 \le n \le L-1$) and a subsystem $B$ containing the remaining chain sites
($L \le n \le N-1$) [see Fig.\ \ref{fig:schematic}(b)]. We obtain the
entanglement entropy $S_e(J,L,N)=-\mathrm{Tr}_A(\rho_A\ln \rho_A)$ by
using the NRG solutions of $H_M$ with $L-1 \le M \le N$ to compute the
reduced density operator for subsystem $A$: $\rho_A = \mathrm{Tr}_B(\rho)$
\cite{Weichselbaum:07,Merker:12,Nghiem:14,Note1}. Here,
$\rho\propto\exp(-H_N/k_B T)$ is the density operator at a thermal energy
scale $k_B T\sim t_N$, chosen to be much smaller than any other energy of
physical interest so that the ground-state entanglement is calculated.
(For $J < J_c$, a tiny magnetic field is introduced to remove a
spurious contribution to $S_e$ from the two-fold ground-state
degeneracy \cite{Pixley:15}.) We also calculate the entanglement entropy
$S_e^{(0)}(L,N)$ for the same partition of the chain but without the
impurity \cite{Note1}. The impurity entanglement entropy, defined as
$\Sei(J,L,N) = S_e(J,L,N) - S_e^{(0)}(L,N)$,
is independent of $N$ provided that $N\gg L$, but shows an alternating term
proportional to $(-1)^L$ that decays only slowly with increasing $L$
\cite{Sorensen:07,L=0}. We therefore focus on a smoothed three-point average
$\Sei(J,L)=\lim_{N\gg L}[\Sei(J,L\!-\!1,N)+2\Sei(J,L,N)+\Sei(J,L\!+\!1,N)]/4$
\cite{Note1}.

To find $S_e$ as a function of physical distance $R$ from the impurity, we
note that site $n$ of the Wilson chain is associated with a single-electron
wave function $\psi_n(r')$ that has its greatest radial probability density
at radius $r'_n \simeq c\L^{n/2}/k_F$, where $k_F$ is the Fermi wave vector
and $c$ is a dimensionless constant of order unity \cite{Krishna-murthy:80.I}
[see Fig.\ \ref{fig:schematic}(a)]. In the physical limit $N\to\infty$ and
$\Lambda\to 1$, $\psi_n(r')$ approaches a radial delta function. Even for
$\Lambda>1$, we expect the smoothed entanglement entropy $\Sei(J,L)$ to
reasonably approximate its continuum counterpart $\Sei(J,R=c\Lambda^{L/2}/k_F)$.
We present results obtained using discretization parameter $\L=3$, retaining
up to 600 many-body eigenstates after each NRG iteration to reach a Wilson
chain of $N=161$ sites. We employ the conventional NRG value
$c=2\L^{1/2}/(\L+1)$ and work in units where $D=\hbar=k_B=g\mu_B=1$
\cite{Note1}.

\begin{figure}
\centering
\includegraphics{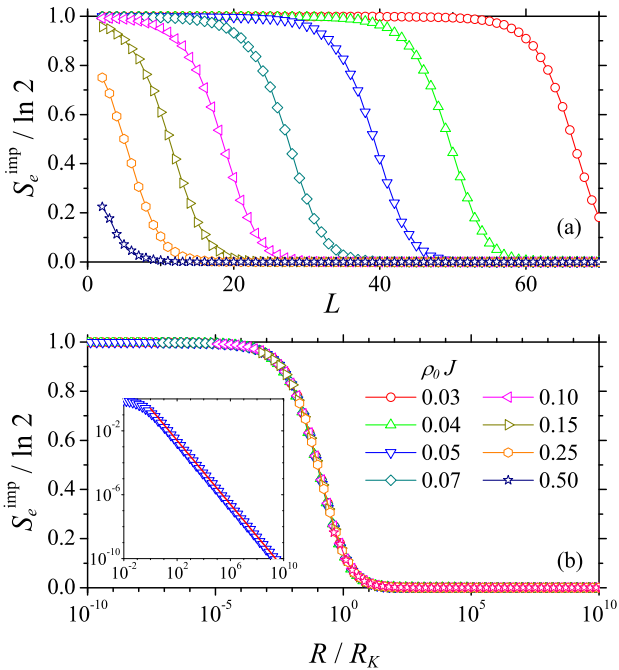}
\caption{\label{fig:r=0}
(a) Impurity entanglement entropy $\Sei$ vs Wilson chain partition
size $L$ for a metallic host ($r=0$) and different Kondo couplings $J$ labeled
in the legend of (b). Lines are guides to the eye.
(b) Data from (a) replotted as $\Sei$ vs $R/R_K$, where $R=c\L^{L/2}/k_F$ and
$R_K=1/(k_F T_K)$ with $T_K$ being the Kondo temperature extracted from the
magnetic susceptibility \cite{Note1}. The collapse of data for different $J$
values points to a one-parameter scaling $\Sei(J,R) = f_0(R/R_K)$.
Inset: Data from main panel for $\rho_0 J = 0.05$ replotted on a log-log scale
showing an $(R/R_K)^{-1}$ tail (fitted line) for $R\gg R_K$.}
\end{figure}

\emph{Results for a metallic host.}
First we consider the conventional Kondo model described by band exponent
$r=0$. Figure \ref{fig:r=0}(a) plots the impurity entanglement entropy
$\Sei$ vs $L$ for eight values of the Kondo coupling $J$. For all but the
largest $J$ values, $\Sei$ starts for small $L$ at the value $\ln 2$
indicative of a singlet formed between (i) a spin $\half$ arising from an
impurity that is negligibly screened by electrons occupying Wilson sites $n<L$,
and (ii) a net spin $\half$ representing the part of the Kondo screening cloud
residing on Wilson chain sites $n\ge L$. For large $L$, $\Sei$ approaches zero
from above, indicating that the impurity is being Kondo-screened almost
entirely by electrons within subsystem $A$, leaving an entanglement with
subsystem $B$ no greater than in the absence of the impurity.

It is natural to associate the crossover from $\Sei\simeq \ln 2$ to $\Sei=0$
with the renormalization-group (RG) flow from weak to strong coupling, known
from much previous work \cite{Hewson:93} to be characterized by a single energy
scale $T_K$. Accordingly, the entanglement is believed \cite{Sorensen:07} to
have just one length scale $R_K\simeq 1 / (k_F T_K)$.
Figure \ref{fig:r=0}(b) replots data from Fig.\ \ref{fig:r=0}(a) as $\Sei$
vs $R/R_K$, revealing an excellent collapse of results for different $J$ and
pointing to the existence of a universal scaling $\Sei(J,R) = f_0(R/R_K)$. For
$R/R_K\gg 1$, $\Sei$ decays like $(R/R_K)^{-1}$ [see inset to Fig.\
\ref{fig:r=0}(b)], consistent with studies of spin chains \cite{Sorensen:07}
and a resonant-level model \cite{Saleur:13}.

\begin{figure}
\centering
\includegraphics{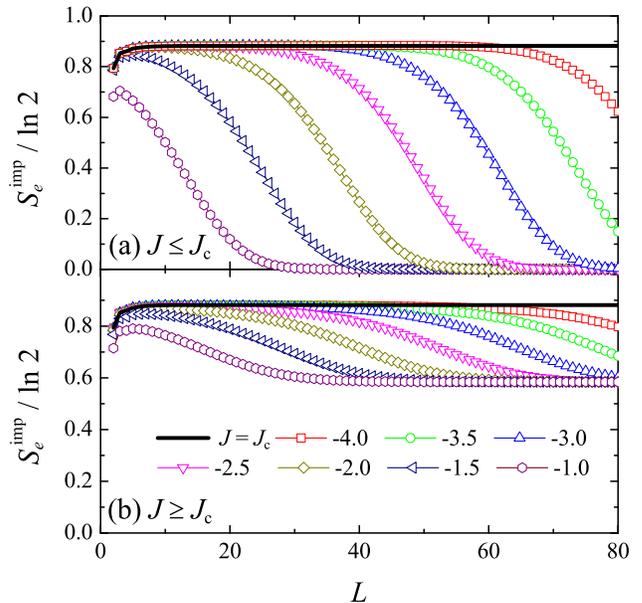}
\caption{\label{fig:r=0.4}
Impurity entanglement entropy $\Sei$ vs Wilson chain partition size
$L$ for a pseudogap Kondo model with band exponent $r=0.4$. Symbols plot data
for (a) $J=(1-10^x)J_c$, and (b) $J=(1+10^x)J_c$, with values of $x$ shown in
the legend. Thick lines show the critical case $J=J_c$.
}
\end{figure}

\begin{figure}
\centering
\includegraphics{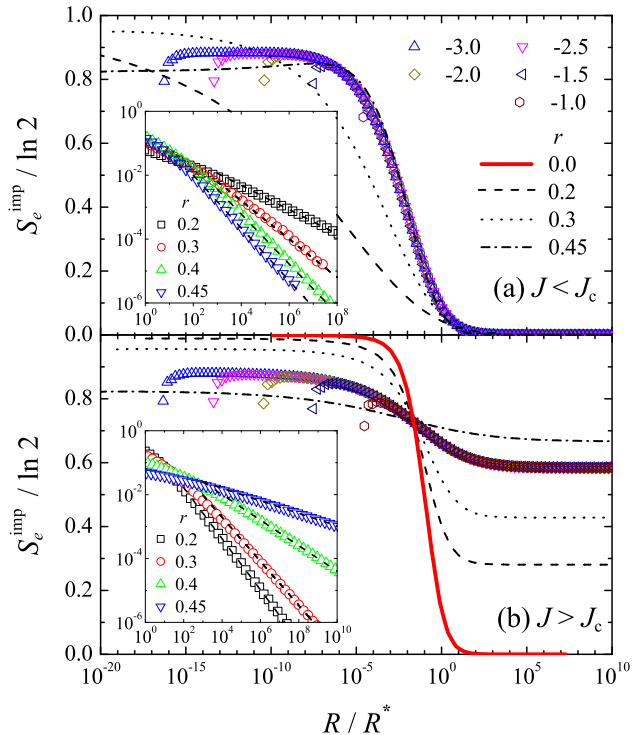}
\caption{\label{fig:r>0}
Data from Fig. \ref{fig:r=0.4} replotted vs $R/R^*$, where
$R\propto\L^{L/2}/k_F$ and $R^*=1/(k_F T^*)$ with $T^*$ being a crossover
temperature extracted from the magnetic susceptibility \cite{Note1}. Symbols
plot data for (a) $J=(1-10^x)J_c$, and (b) $J=(1+10^x)J_c$ with values of $x$
labeled in the legend. Lines show fits to data points (not shown) obtained
for other values of $r$. Insets: Log-log plots of large-$R$ data for
$\Sei(J,R)-\Sei(J,\infty)$ vs $R/R^*$, calculated for a single Kondo coupling
(a) $J < J_c$, (b) $J > J_c$ at each of four different band exponents $r>0$,
with power-law fits (dashed lines).
}
\end{figure}

\emph{Results for pseudogapped hosts.}
Our main interest is in the entanglement near the Kondo-destruction QCPs that
occur for semimetallic densities of states described by exponents $0<r<\half$.
Figure \ref{fig:r=0.4} illustrates for $r=0.4$ the variation of $\Sei$ with
Wilson chain partition size $L$ for values of $J$ close to $J_c$. In the
local-moment phase [Fig.\ \ref{fig:r=0.4}(a)], $\Sei$ initially rises with
increasing $L$ to reach a plateau maximum, only to fall toward zero for larger
partition sizes. These data show that even though the impurity spin
asymptotically decouples from the band, the impurity induces additional
entanglement for finite values of $L$---or equivalently, at finite energies
$\simeq \pm D\L^{-L/2}$---manifesting a \emph{dynamical} Kondo effect.

In the Kondo phase, too, $\Sei$ initially rises with increasing $L$ to reach
the same plateau maximum as for $J<J_c$, before decreasing for larger $L$
values [Fig.\ \ref{fig:r=0.4}(b)]. Here, however, the impurity induces an
additional entanglement that remains nonzero as $L\to\infty$. This is
consistent with the nonvanishing $T\to 0$ limits of both the impurity entropy
and the effective magnetic moment, which suggest that the impurity degree of
freedom is only partially screened in the pseudogap Kondo phase
\cite{Gonzalez-Buxton:98}.

Figure \ref{fig:r=0.4} also shows that in either phase, $\Sei$ remains near its
initial plateau to larger values of $L$ the closer $J$ approaches $J_c$. We are
thus led to one of our principal conclusions: At the QCP [thick lines in Figs.\
\ref{fig:r=0.4}(a) and \ref{fig:r=0.4}(b)], the \emph{entire} conduction band is
maximally entangled with the impurity, i.e., the ground state has long-range,
scale-invariant entanglement.

The preceding picture implies that the eventual decrease in $\Sei$ vs $L$ seen
for $J\ne J_c$ reflects the RG flow away from the pseudogap
Kondo QCP, a flow characterized by a crossover temperatures scale
$T^*\sim |J-J_c|^{\nu}$, where $\nu(r)$ is the correlation-length exponent
\cite{Ingersent:02}. Following the same reasoning as was applied for a
metallic ($r=0$) host, we expect $T^*$ to be associated with a length scale
$R^*=1/(k_F T^*)$. Figure \ref{fig:r>0} replots the $r=0.4$ data from
Fig.\ \ref{fig:r=0.4} as $\Sei$ vs $R/R^*$ using values of $T^*$ extracted from
the magnetic susceptibility \cite{Gonzalez-Buxton:98,Note1}. The scaling
collapse of data for different $J$ is of a similar quality to that for $r=0$
[see Fig.\ \ref{fig:r=0}(b)]. This provides strong evidence for the existence
of scaling functions $f_r^{\pm}$ such that
\begin{equation}
\label{eq:scaling:r>0}
\Sei(J,R) = f_r^{\pm}(R/R^*)  \quad \text{for} \quad J\gtrless J_c.
\end{equation}
Significant departures from scaling are seen only for the smallest values of
$R$ (corresponding to the smallest $L$ in Fig.\ \ref{fig:r=0.4}), and
can be attributed to the NRG discretization.
Figure \ref{fig:r>0} also plots fitting curves from similar data collapses
for band exponents $r=0.2$, $r=0.3$ and $r=0.45$ \cite{Note2}, as well as
[in panel (b)] the metallic case $r=0$.

Whereas in the local-moment phase $\Sei\to 0$ for $R/R^*\to\infty$, in the
Kondo phase $\Sei$ approaches for $R/R^*\gg 1$ a value that is
well-approximated by $\Sei\simeq\frac{3}{2}r\ln 2$ \cite{Note1}. Insets in
Fig.\ \ref{fig:r>0} show that in either phase, the impurity entanglement
entropy has a power-law tail
\begin{equation}
\Sei(J,R)-\Sei(J,\infty)\propto(R/R^*)^{-\alpha} \quad \text{for } R\gg R^*.
\end{equation}
Fitted exponents are consistent with $\alpha = 2r$ for $J < J_c$ and
$\alpha = \min(1\!-\!r,\ 2\!-\!4r)$ for $J>J_c$, values that
represent twice the dimension of the leading irrelevant operator at the
local-moment and Kondo fixed points, respectively \cite{Note1}. This
observation is consistent with the interpretation that the power-law tails
are associated with the RG flow toward a stable fixed point.

\emph{Discussion.}
We have determined the spatial structure of entanglement entropy in two types
of quantum impurity models. Our work demonstrates that the impurity entanglement
entropy for a system partitioned at radius $R$ around a Kondo impurity depends
only on $R$ divided by $R^* \propto 1/T^*$, where $T^*$ is a many-body scale.
In the conventional case of a metallic host, $T^*$ is the Kondo temperature,
whereas $T^*$ vanishes like $|J-J_c|^{\nu}$ on approach to the
Kondo-destruction critical point in a pseudogapped host. The impurity
entanglement entropy is both scale invariant and long ranged at this
interacting critical point, while away from criticality it falls off like
$\sim(R/R^*)^{-\alpha}$ for $R\gg R^*$. We deduce that the \emph{total\/}
entanglement entropy goes like $S_e(J,R)\simeq b\log R + f_r^{\pm}(R/R^*)$.
Our conclusions have been reached for a model [Eqs.\ \eqref{eq:Ham} and
\eqref{eq:dos}] exhibiting strict particle-hole symmetry, but we expect
similar conclusions to apply at the asymmetric interacting QCPs that arise
for $0.375 \lesssim r<1$ upon the addition of a potential-scattering term to\
Eq.\ \eqref{eq:Ham} \cite{Gonzalez-Buxton:98}.

Although obtained for a single impurity, our results shed light on the Kondo
lattice model and have implications for interpretation of experiments on
quantum-critical heavy-fermion compounds. First, this work provides insight
into the structure of the ground-state wave function in a Kondo-destroyed
phase. Even though static screening is suppressed, a dynamical Kondo effect
still produces entanglement extending over a length scale that diverges on
approach to the Kondo phase boundary. Thus, the ground state of the Kondo
lattice cannot be adequately described merely in terms of a static slave-boson
amplitude; dynamical effects must be taken into account. Second, our findings
suggest that the Kondo-breakdown QCP relevant to heavy-fermion metals is
accompanied by long-range entanglement between all local moments and the
entire conduction band. We believe that this scale-invariant entanglement is
intimately associated with the reconstruction of the critical Fermi surface.

\emph{Acknowledgments.}
We thank Andreas Ludwig and Andrew Mitchell for useful discussions.
This work was supported in part by NSF Grant No.\ DMR-1508122 (C.W., T.C.,
and K.I.), by the DFG within the CRC 1238 (Project C03), and by the Laboratory for Physical Sciences (J.H.P). The work of K.I.\ was performed in
part at the Aspen Center for Physics, which is supported by NSF Grant
No.\ PHY-1066293.

\clearpage
\widetext

\newpage
\onecolumngrid
\setcounter{figure}{0}
\makeatletter
\renewcommand{\thefigure}{S\@arabic\c@figure}
\setcounter{equation}{0} \makeatletter
\renewcommand \theequation{S\@arabic\c@equation}

\begin{center}
\textbf{\large Supplemental Material for ``Long-range entanglement near a
Kondo-destruction quantum critical point''} \\[2.5ex]

Christopher Wagner$^{1}$, Tathagata Chowdhury$^{1,2}$, J.\ H.\ Pixley$^{3,4}$,
and Kevin Ingersent$^{1}$ \\[1ex]

{\small
${}^{1}$\textit{Department of Physics, University of Florida, Gainesville,
Florida 32611-8440, USA} \\
${}^{2}$\textit{Institut f\"{u}r Theoretische Physik, Universit\"{a}t zus
K\"{o}ln, Z\"{u}lpicher Strasse 77a, 50937 K\"{o}ln, Germany} \\
${}^{3}$\textit{Department of Physics and Astronomy, Center for Materials
Theory, Rutgers University, Piscataway, NJ 08854 USA} \\
${}^{4}$\textit{Condensed Matter Theory Center and the Joint Quantum
Institute, \\ Department of Physics, University of Maryland, College Park,
Maryland 20742-4111, USA} \\
(Dated: \today)}
\end{center}

This document summarizes technical methods and presents results beyond those
contained in the main paper.
Section \ref{sec:WC_entanglement} analyzes the Wilson chain---the discretized
representation of the conduction band used in numerical renormalization-group
(NRG) calculations---in the absence of any impurity degree of freedom. Data are
presented for the dependence of the entanglement entropy $S_e$ on the overall
chain length, the position of the cut across which the entanglement is
computed, the exponent $r$ entering the density of states
\begin{equation}
\rho(\veps) = \rho_0 |\veps|^r \Theta(D-|\veps|),
\label{DOS}
\end{equation}
and the NRG discretization parameter $\Lambda$.
Section \ref{sec:Kondo_entanglement} addresses the many-body Kondo problem
created by coupling a Wilson chain to a spin-$\half$ impurity. A
description of the method that we use to calculate the entanglement entropy is
followed by details of the entanglement results presented in the main text.

\section{Entanglement within the Wilson chain}
\label{sec:WC_entanglement}

This section focuses on the entanglement properties of isolated Wilson chains
(without any coupling to an impurity degree of freedom). The quantity of interest
is the entanglement entropy $S_e(L)$ for a chain of length $N$ sites that is
split into subsystem $A$ comprising sites $0$ through $L-1$ and subsystem $B$
containing sites $L$ through $N-1$. Although the notation suggests that $S_e$ is a
function of $L$ alone, it must be emphasized that in fact it also depends on $N$,
$r$, and $\Lambda$.

\subsection{The Wilson chain}
\label{subsec:WC}

The NRG method \cite{Wilson:75-2,Bulla:08-2} uses a discretization parameter
$\Lambda>1$ to divide a conduction band having single-particle energies $\veps$
ranging from $-D$ to $D$ into an infinite set of logarithmic bins spanning
$D\Lambda^{-(m+1)} < \pm \veps \le D\Lambda^{-m}$ for $m = 0$, $1$, $2$,
$\ldots$. Within each bin, the band is approximated by a single representative
state, namely, the linear combination of the original band states that couples
to the impurity. The band Hamiltonian is then mapped via the Lanczos
method onto a tight-binding Hamiltonian for a semi-infinite ``Wilson chain'' of
sites labeled $n = 0$, $1$, $2$, $\ldots$, coupled to the impurity via site
$0$ only:
\begin{equation}
H \longrightarrow H_{\imp}\bigl[f_{0\sigma}^{\pdag}, f_{0\sigma}^{\dag}\bigr]
+ \sum_{n=1}^{\infty} \sum_{\sigma} t_n
\left( f_{n\sigma}^{\dag} \, f_{n-1,\sigma}^{\pdag} + \text{H.c.} \right) ,
\label{WC}
\end{equation}
where $t_n \sim D\Lambda^{-n/2}$ for $n\gg 1$ \cite{on-site-2}.
The reader is referred to Ref.\ \onlinecite{Gonzalez-Buxton:98} for details of
the calculation of the coefficients $t_n$ and their large-$n$ asymptotics for
the power-law density of states specified in Eq.\ \eqref{DOS}.

The discretization-induced separation of energy scales $t_n$ allows controlled
approximation of the low-energy states of the full Hamiltonian $H$ through
iterative solution of finite-chain Hamiltonians
\begin{align}
\label{H_M}
H_M &= H_{\imp}\bigl[f_{0\sigma}^{\pdag}, f_{0\sigma}^{\dag}\bigr]
  + H^{\chain}_M, \\
\label{H_M^chain}
H^{\chain}_M &= \sum_{n=1}^{M-1} \sum_{\sigma} t_n \left( f_{n\sigma}^{\dag} \,
f_{n-1,\sigma}^{\pdag} + \text{H.c.} \right),
\end{align}
with $M = 1$, $2$, $\ldots$, $N$. Here, $N$ is chosen to be sufficiently
large that $t_N$ (the largest energy scale of the part of the semi-infinite
chain that is omitted from $H_N$) is much smaller than all energy scales of
physical interest.

\begin{figure}
\includegraphics[width=0.5\linewidth]{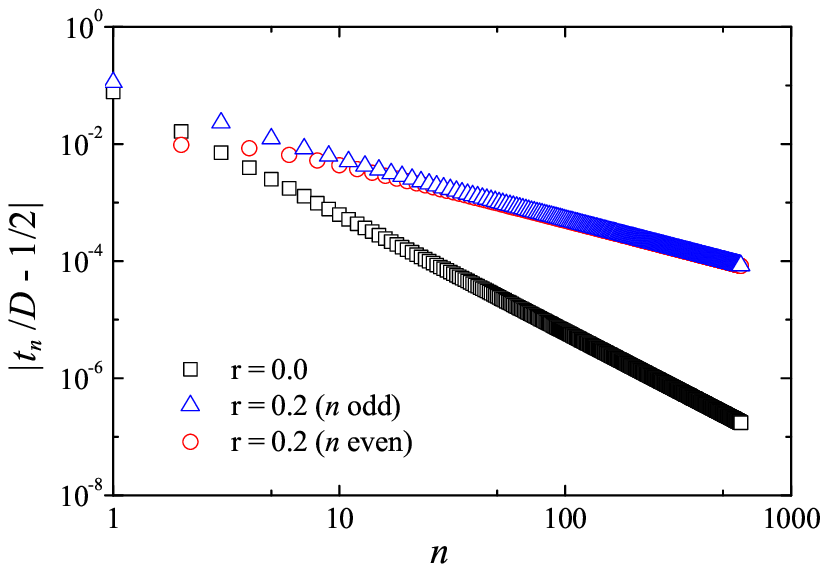}
\caption{\label{fig:t_n}%
Tight-binding hopping parameters plotted as $|t_n/D-\half|$ vs $n$
for the exact Lanczos mapping of a conduction band with a density of states
given by Eq.\ \eqref{DOS} with $r=0$ (squares) and $r=0.2$ (triangles and
circles for odd and even $n$, respectively). Only for $n\to\infty$ does
$t_n$ approach its uniform value $t_n = \half$ for a standard
tight-binding chain.
}
\end{figure}

The Wilson chain hopping coefficients converge for $\Lambda \rightarrow 1$ to
those for the exact Lanczos mapping of the continuum (``$\Lambda=1$'') Kondo
model. For example, in the case of a metallic density of states
[Eq.\ \eqref{DOS} with $r=0$], $t_n$ decreases monotonically from
$t_1 \approx 0.57 D$ toward $t_{\infty}=D/2$. The log-log plot in Fig.\
\ref{fig:t_n} reveals an exponential decay of $t_n/D-\half$ with
increasing $n$. This pattern distinguishes the exact tight-binding formulation
of the Kondo model from a standard tight-binding (STB) chain corresponding to
Eq.\ \eqref{H_M^chain} with $t_n = D/2$. The effect of this
difference on the entanglement entropy will be discussed below. Figure
\ref{fig:t_n} also plots $|t_n/D-\half|$ vs $n$ for the pseudogapped
case $r=0.2$. Here $t_n$ for $n$ odd (even) approaches $D/2$ from above
(below).

\subsection{Calculation of the entanglement entropy}
\label{subsec:WC_calculation}

The entanglement entropy of an isolated Wilson chain (without any impurity
coupled to its end) can be computed within the NRG approach in the same manner
as the corresponding quantity for the full Kondo problem (see Sec.\
\ref{subsec:NRG_calculation}). However, the quadratic nature of $H_M^{\chain}$
allows use of a simpler method introduced in Ref.\ \onlinecite{Peschel:03},
which was used to obtain the results presented in Sec.\ \ref{subsec:WC_results}.

The formalism begins with a general fermionic tight-binding Hamiltonian
\begin{align}
H = \sum_{m,n} t_{mn} c_m^{\dag} c_n^{\pdag}
\end{align}
having hopping $t_{mn}$ between sites $m$ and $n$ of a finite lattice. 
We split the system into subsystems $A$ and $B$, reserving labels $i$ and
$j$ for sites within $A$. In any many-particle eigenstate $|\Psi\rangle$
of $H$, the single-particle correlation function for subsystem $A$ can be
written
\begin{equation}
C_{ij} = \langle \Psi |c_i^{\dag} c_j ^{\pdag}| \Psi \rangle
= \text{Tr}_A \bigl( \rho_A c_i^{\dag} c_j^{\pdag} \bigr),
\label{C_ij}
\end{equation}
where $\rho_A = \text{Tr}_B (|\Psi\rangle\langle\Psi|)$ is the reduced
density operator for subsystem $A$.
Given the quadratic form of $H$, higher correlation functions within
$A$ must factorize according to Wick's theorem, and there must
exist a Hermitian operator
\begin{equation}
\mathcal{H}_A = \sum_{i,j} \tilde{H}_{ij} \, c_i^{\dag} c_j^{\pdag}
\label{tildeH}
\end{equation}
such that
\begin{equation}
\rho_A = Z_A^{-1} e^{-\mathcal{H_A}}, \qquad
Z_A = \text{Tr}_A \bigl( e^{-\mathcal{H_A}} \bigr) .
\label{rho_A}
\end{equation}

The matrix $\tilde{H}_{ij}$ has a set of eigenvalues $\tveps_k$
and orthonormal eigenvectors $\mathbf{v}_k$ with components $v_{ik}$ that can
be used to define new fermionic operators
\begin{equation}
a_k = \sum_i v_{ik}^* \, c_i
\quad \longleftrightarrow \quad
c_i = \sum_k v_{ik} \, a_k
\label{a_k}
\end{equation}
such that
\begin{equation}
\mathcal{H}_A = \sum_k \tveps_k \, a_k^{\dag} a_k^{\pdag} , \qquad
\rho_A = Z_A^{-1} \exp \biggl( -\sum_k \tveps_k \, a_k^{\dag}
a_k^{\pdag} \biggr) , \qquad
Z_A = \prod_k \bigl (1 + e^{-\tveps_k} \bigr).
\label{tildeH:diag}
\end{equation}

Substituting Eq.\ \eqref{a_k} into Eq.\ \eqref{tildeH} yields
\begin{equation}
\tilde{H}_{ij} = \sum_k \tveps_k \, v_{ik} v_{jk}^* ,
\label{tildeH:1}
\end{equation}
while substituting Eqs.\ \eqref{rho_A}--\eqref{tildeH:diag} into
Eq.\ \eqref{C_ij} gives
\begin{equation}
C_{ij} = \sum_k \frac{v_{ik}^* v_{jk}}{e^{\tveps_k} + 1}.
\label{C_ij:1}
\end{equation}
Comparison of Eqs.\ \eqref{tildeH:1} and \eqref{C_ij:1} leads to the
conclusion that matrices $\tilde{H}$ and $C^T$ (the transpose of $C$)
are diagonalized by the same similarity transformation. We therefore
deduce that $C^T$ (and hence $C$) has eigenvalues 
\begin{align}
\gamma_k = \frac{1}{1 + e^{\tveps_k}} \quad \longleftrightarrow \quad
\tveps_k = \ln \left( 1 - \gamma_k \right) - \ln \gamma_k.
\end{align}

The entanglement entropy for the partition of the system into subsystems
$A$ and $B$ is
\begin{align}
S_e
&= - \text{Tr}_A ( \rho_A \ln \rho_A ) \notag \\
&= \text{Tr}_A  ( \rho_A \ln Z_A )
+ \text{Tr}_A \left[ Z_A^{-1} \exp \biggl(-\sum_k \tveps_k \, a_k^{\dag}
a_k^{\pdag} \biggr) \biggl( \sum_{k'} \tveps_{k'} \, a_{k'}^{\dag}
a_{k'}^{\pdag} \biggr) \right] \notag \\
&= \ln Z_A + \prod_k \frac{\tveps_k}{e^{\tveps_k} + 1}
\: = \: \sum_k \ln \bigl( 1 + e^{-\tveps_k} \bigr)
+ \sum_k \frac{\tveps_k}{e^{\tveps_k} + 1} \notag \\
&= \sum_k \ln \biggl( 1 + \frac{\gamma_k}{1-\gamma_k} \biggr)
+ \sum_k \gamma_k \bigl[ \ln \bigl( 1 - \gamma_k \bigr)
- \ln \gamma_k \bigr] \notag \\
&= - \sum_k \bigl[ \gamma_k \ln \gamma_k + (1 - \gamma_k)
\ln (1 - \gamma_k) \bigr].
\label{S_e:final}
\end{align}
Equation \eqref{S_e:final} provides a computationally fast and accurate
method for calculating the entanglement of a spinless Wilson chain or a
spinless standard tight-binding chain from the eigenvalues of its
single-particle correlation function.
To obtain the impurity contribution of the entanglement entropy of the
Kondo problem, we subtract twice the entanglement entropy of the spinless
chain.

\subsection{Systematics of the Wilson chain entanglement entropy}
\label{subsec:WC_results}

The entanglement entropy of the Wilson chain exhibits even-odd alternation
with increasing size $L$ of partition $A$. Such an alternation is present
for a standard tight-binding chain, but it becomes more pronounced with
increasing $\Lambda>1$ and/or increasing $|r|$. To filter out this
alternation, which is a finite-size effect of little interest for our
purposes, we consider a three-point average
\begin{equation}
\Sea(L) = \frac{1}{4} \bigl[S_e(L-1) + 2 S_e(L) + S_e(L+ 1)\bigr].
\end{equation}

This section considers first the case of a metallic band with a density of
states described by Eq.\ \eqref{DOS} with $r=0$. We identify a range of $L$
values over which $\Sea$ differs negligibly from the universal
dependence exhibited by a standard tight-binding (STB) chain, and describe
deviations found for small and large values of $L$. We then turn to the
effects of varying the band exponent $r$ entering Eq.\ \eqref{DOS}.

\begin{figure}
\includegraphics[width=1.0\linewidth]{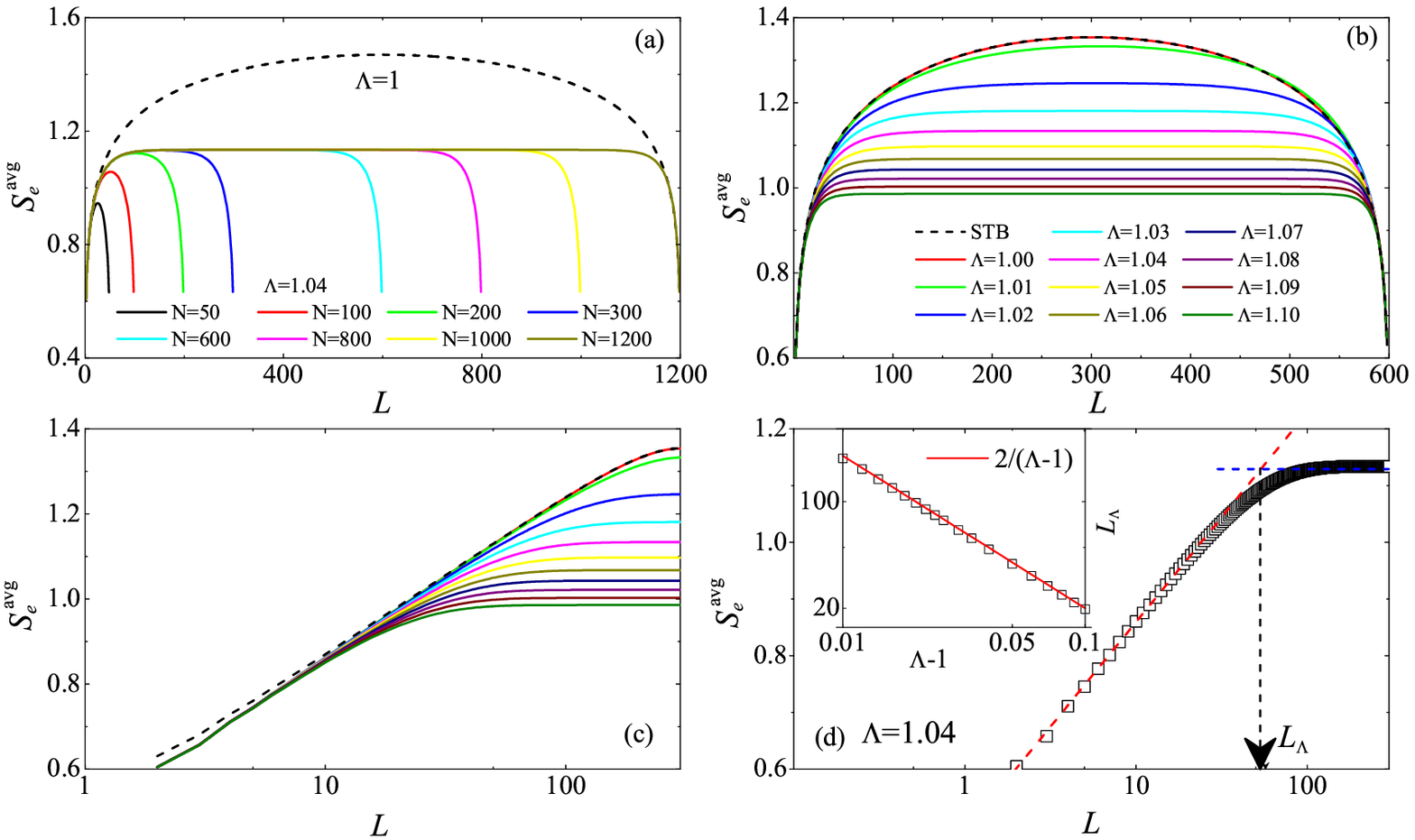}
\caption{\label{fig:Se-vs-L:r=0}%
Wilson chain entanglement entropy $\Sea$ vs partition size $L$ for
a metallic density of states described by Eq.\ \eqref{DOS} with $r=0$.
(a) Data for discretization parameter $\Lambda = 1.04$ with different chain
lengths $N$ (solid lines), and for $\Lambda = 1$, $N = 1200$ (dashed  line).
(b) Data for $N = 600$ with different values of $\Lambda$ (solid lines).
Also shown (dashed line) is $\Sea$ vs partition size $L$ for a
600-site standard tight-binding (STB) chain.
(c) Data from (b) replotted vs $\log L$.
(d) Data for $N=600$, $\Lambda = 1.04$ showing the definition of a
partition size $L_\Lambda$ characterizing the crossover from a regime
$\Sea \sim \log L$ for $L\ll L_\Lambda$  (red dashed line)
to a regime $\Sea \simeq \Sem$ for $L\gg L_\Lambda$ (blue dashed line).
Inset: $L_\Lambda$ (calculated for $N=1200$) vs $\Lambda-1$ is well
approximated by $L_\Lambda = 2/(\Lambda-1)$ (red line).
}
\end{figure}

\emph{Results for $r=0$}:
Fig.\ \ref{fig:Se-vs-L:r=0}(a) shows the average entanglement entropy
$\Sea$ vs partition length $L$ for a representative case $\Lambda=1.04$ and
for various chain lengths $N$ specified in the legend. $\Sea(L)$ is almost
(but not quite) symmetric with respect to reflection about $L=N/2$ and peaks
very close to $L=N/2$. The value $\Sem\simeq\Sea(N/2)$ initially increases
with increasing chain length $N$, but eventually saturates as a wide plateau
forms in $\Sea(L)$. No such plateau is observed in the data for $\Lambda=1$
and $N=1200$ (plotted with dashed lines).

Figure \ref{fig:Se-vs-L:r=0}(b) plots (solid lines) $\Sea$ vs $L$
for a fixed chain length $N=600$ and different values of the discretization
parameter on the range $1 \leq \Lambda \leq 1.1$. Also shown (dashed line)
are the corresponding data for a standard tight-binding (STB) chain with
$t_n=D/2$. The STB curve is exactly symmetric about $L=N/2$, while those for
Wilson chains are slightly asymmetric. Curves for $\Lambda>1$ exhibit a
plateau similar to that seen in Fig.\ \ref{fig:Se-vs-L:r=0}(a).
As $\Lambda$ is increased, the plateau value $\Sem$ decreases and is
reached at smaller values of $L$.

For the fermionic STB chain with constant hopping coefficients between
nearest neighbors, the entanglement entropy in the limit $L\ll N$ is
equal to that of a critical conformal field theory (CFT)
\cite{Calabrese:04-2}. For a finite system with open boundary conditions,
\begin{align}
\Sea
&= \frac{c}{6}\ln \Biggl( \frac{N}{\pi} \sin  \frac{\pi L}{N} \Biggr) + b
  \notag \\
& \simeq \: \frac{c}{6}\ln L + b \; \; \; \; \text{for $L \ll N/2$},
\label{Se_CFT}
\end{align}
where $c$ is the central charge of the CFT and $b$ the boundary entanglement.
For a spinless chain, the left-moving and right-moving fermions each carry a
charge of $c=1/2$, so the chain overall is described by $c=1$.
Fig.\ \ref{fig:Se-vs-L:r=0}(c) replots the data for $L\le N/2$ from
Fig.\ \ref{fig:Se-vs-L:r=0}(b) as $\Sea$ vs $\log L$.
The Wilson chain results (solid lines) can be well approximated by
\begin{equation}
\Sea = \begin{cases}
{\displaystyle\frac{c}{6}} \ln L + b &
  \text{for $10 \lesssim L \ll L_{\Lambda}$}, \\[2ex]
{\displaystyle\frac{c}{6}} \ln L_{\Lambda} + b \equiv \Sem &
  \text{for $L_{\Lambda} \ll L \le N/2$}.
\end{cases}
\label{Sea:form}
\end{equation}
Here, $c$ and $b$ are independent of $\Lambda$ and, when extrapolated to the
infinite-size limit $1/N\to 0$, are numerically indistinguishable from their
respective STB-chain values: $c=1$ and $b\simeq 0.478$.
For $L\lesssim 10$, all Wilson-chain data coincide but clearly differ
from those for the STB chain (dashed line), while the STB and $\Lambda=1$ Wilson
chain entanglement entropies converge for $L\gg 10$.
This is unsurprising given the approach with increasing $n$ of the $\Lambda=1$
Wilson chain hopping coefficients $t_n$ to the STB value $t_n = D/2$ (see
Fig.\ \ref{fig:t_n}).

The scale $L_{\Lambda}$ is the focus of Fig.\ \ref{fig:Se-vs-L:r=0}(d).
The main panel shows how $L_{\Lambda}$ can be defined as the horizontal
coordinate of the intercept between the small-$L$ and large-$L$ asymptotes
defined in Eq.\ \eqref{Sea:form}, i.e.,
$L_{\Lambda} = \exp[(6/c)(\Sem-b)]$.
The inset of Fig.\ \ref{fig:Se-vs-L:r=0}(d) plots the variation of
$L_{\Lambda}$ with $\Lambda$ (data points), demonstrating that for
$\Lambda \lesssim 1.1$, the scale is well-described by the empirical
relation $L_{\Lambda} = 2/(\Lambda -1)$ (line). NRG many-body calculations
are typically performed using a discretization parameter on the range
$1.5 \le \Lambda \le 3$ chosen to balance discretization errors against
truncation errors. In all such cases, $L_\Lambda \simeq 1$, so
$\Sea(L)\simeq\Sem$ is almost independent of $L$.

\begin{figure}[t]
\centering
\includegraphics[width=0.8\linewidth]{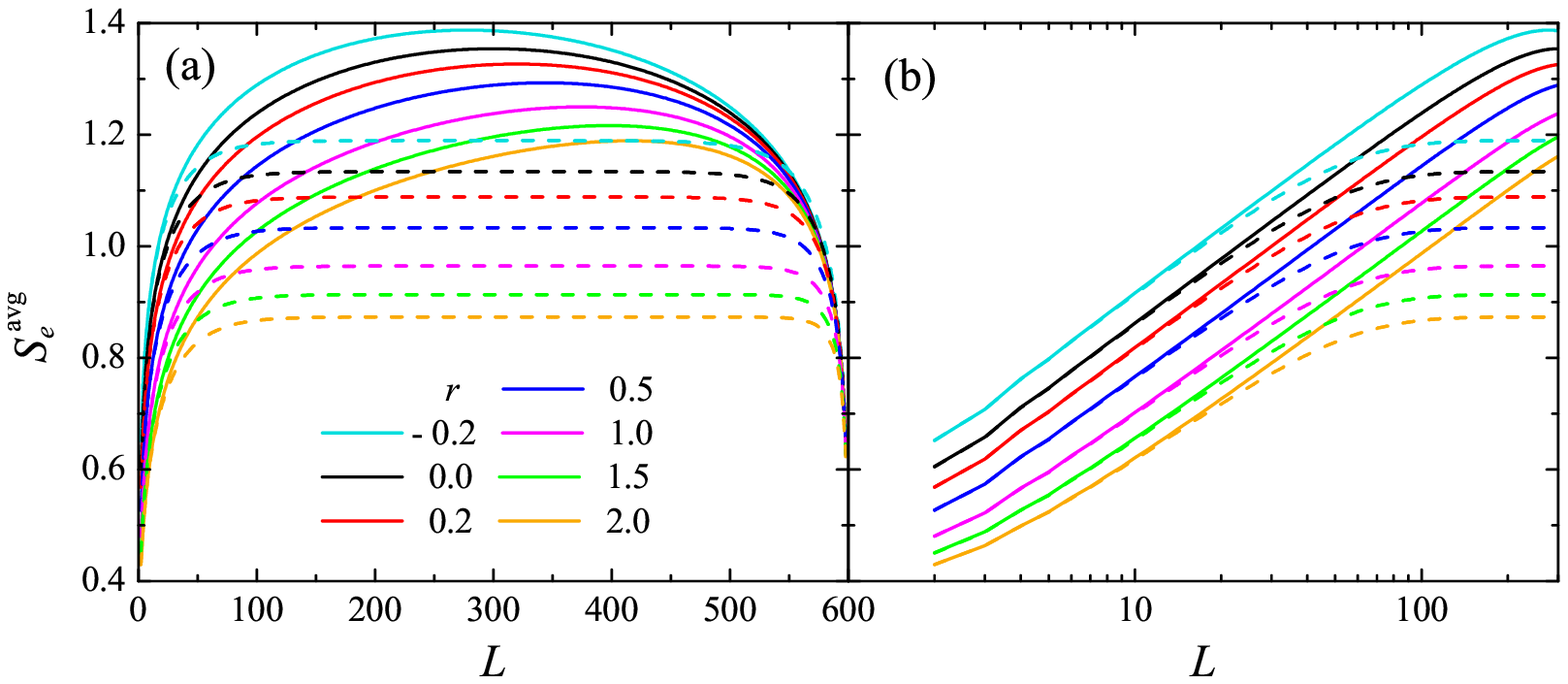}
\caption{\label{fig:Se-vs-L:r>0}%
(a) $\Sea$ vs $L$ for Wilson chains of length $N=600$ with different band
exponents $r$, for discretization parameters $\Lambda=1.0$ (solid lines)
and $\Lambda=1.04$ (dashed lines).
(b) Data for $L\le N/2$ replotted on a logarithmic $L$ scale.
}
\end{figure}

\begin{figure}[t]
\centering
\includegraphics[width=0.85\linewidth]{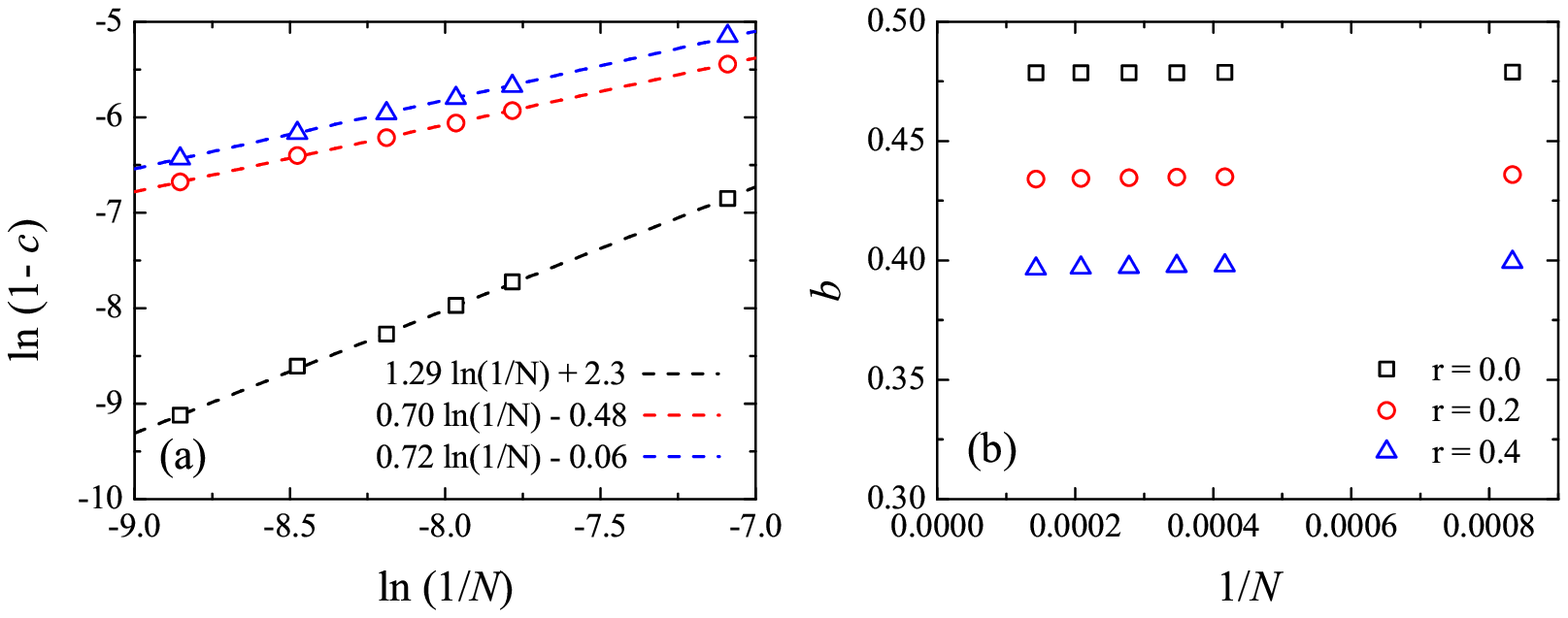}
\caption{\label{fig:c-and-b}%
Fitted coefficients $c$ and $b$ in Eq.\ \eqref{Sea:form} for a Wilson chain
with discretization $\Lambda=1$, and band exponents $r=0$, $0.2$,
and $0.4$. (a) Log-log plot of $1-c$ vs $1/N$, where $N$ is the chain length,
showing apparent convergence to $c=1$ for $1/N\to 0$. (b) $b$ vs $1/N$.
}
\end{figure}

\emph{Results for $r \ne 0$}:
Figure \ref{fig:Se-vs-L:r>0} plots $\Sea$ vs $L$ (panel a) and vs $\log L$
(panel b) for fixed $N=600$, for $\Lambda=1$ (solid lines) and $\Lambda=1.04$
(dashed lines), and for different values of the band exponent $r$ entering
Eq.\ \eqref{DOS} describing metallic ($r=0$), pseudogapped ($r>0$),
and divergent ($r<0$) densities of states.
For $\Lambda=1$, the main effect of increasing $|r|$ is a progressive increase
in the asymmetry of $\Sea(L)$ about $L=N/2$. As $r$ increases (decreases) from
zero, the peak in $\Sea(L)$ moves right (left) from $L\simeq N/2$. 
For $10 \lesssim L \ll N/2$, the entanglement entropy is still described by
Eq.\ \eqref{Se_CFT}, as can be seen from Fig.\ \ref{fig:Se-vs-L:r>0}(b).
For $\Lambda=1.04$ (dashed lines in Fig.\ \ref{fig:Se-vs-L:r>0}), the
entanglement entropy for all $r$ values remains consistent with
Eq.\ \eqref{Sea:form}, where $L_{\Lambda}$ is independent of $r$ and the
value of $\Sem$ tracks the $r$ dependence of $b$, i.e.,
$\Sem(r)-\Sem(0) \simeq b(r) - b(0)$.

Figure \ref{fig:c-and-b} plots the variation with inverse chain length
$1/N$ of the fitted values of $c$ and $b$ for $\Lambda = 1$ and $r=0$,
$0.2$, and $0.4$. Table \ref{tab:c-and-b} lists the result of
extrapolating $c$ and $b$ to the long-chain limit $1/N\to 0$, along with
the corresponding values for the STB chain. To within numerical accuracy,
the slope remains $c=1$ independent of $r$, as demonstrated by a log-log
plot of $1-c$ vs $1/N$ [Fig.\ \ref{fig:c-and-b}(a)], whereas the boundary
entanglement $b$ decreases (increases) as $r$ is increased (decreased)
from zero [Figs.\ \ref{fig:Se-vs-L:r>0}(b) and \ref{fig:c-and-b}(b)].

\begin{table}[t]
\begin{tabular}{lll}
\multicolumn{1}{l}{chain}
  & \multicolumn{1}{c}{$c$)}
	  & \multicolumn{1}{c}{$b$} \\ \hline 
STB     & 1.0000(2) &  0.4780(7) \\
$r=0$   & 1.0000(1) &  0.47856(5) \\
$r=0.2$ & 1.0000(1) &  0.43288(4) \\
$r=0.4$ & 1.0000(1) &  0.39500(5)
\end{tabular}
\caption{\label{tab:c-and-b}%
Values of the coefficients $c$ and $b$ defined in Eq.\ \eqref{Sea:form}
for the STB and for Wilson chains with different band exponents $r$.
A number in parentheses denotes the estimated nonsystematic error in
the last digit.}
\end{table}

A density of states of the form of Eq.\ \eqref{DOS} describes free fermions
in one spatial dimension having a dispersion
$\veps\propto |k-k_F|^{1/(1+r)}\text{sgn}\,(k-k_F)$. It is therefore quite
surprising that, apart from a nonuniversal boundary term $b(r)$, the $L$
dependence of $\Sea$ for $10\lesssim L \ll N/2$ is the same for $r=0$ (where
the host system exhibits conformal invariance) and for $r\ne 0$ (where the
space and time axes are manifestly inequivalent).
At present we do not fully understand the physical origin of this result.
However, it suggests that the pseudogap host could in fact possess a
``hidden'' conformal symmetry with a central charge $c=1$ ($c=\half$ each
for left- and right-movers).

\section{Entanglement entropy for the pseudogap Kondo problem}
\label{sec:Kondo_entanglement}

\subsection{Numerical renormalization-group calculation of entanglement entropy}
\label{subsec:NRG_calculation}

To calculate the entanglement entropy, we employ the full density-matrix NRG
approach \cite{Weichselbaum:07-2,Merker:12-2,Nghiem:14-2}. In order to explain our
method, it is necessary first to briefly review aspects of the conventional
NRG approach. If the impurity has $d_i$ internal states (e.g., $d_i = 2$ for a
spin $S_{\imp}=\half$) and each Wilson chain site has $d$ possible states
(e.g., $d = 4$ for the single, spinful conduction band considered in the present
work), then $H_M$ has a basis of dimension $d_i d^M$. Due to the exponential
growth of this dimension with increasing $M$, starting at some NRG iteration $M_0$
(typically $M_0 = 5$), the basis must be truncated to keep within acceptable
bounds the computational time for setting up and diagonalizing
a matrix representation of $H_M$. The many-body eigenstates of $H_M$ are divided
into two sets: the high-energy states $|l,M\rangle$, $1\le l\le n^{\disc}_M$ are
discarded, and only the lowest-energy states $|k,M\rangle$,
$1\le k \le n_M^{\kept}$ are kept to set up the next Hamiltonian $H_{M+1}$,
which then has a truncated basis of dimension
$d n_M^{\kept} \equiv n_{M+1}^{\kept} + n_{M+1}^{\disc}$.

The full density-matrix NRG approach is constructed around a complete basis of
dimension $d_i d^N$ for the longest Wilson chain (corresponding to $M=N$).
This basis, introduced by Anders and Schiller \cite{Anders:05-2,Anders:06-2},
comprises all states of the form
$|l,e,M\rangle=|l,M\rangle\otimes|e,M\rangle$ where $M$ ranges from $M_0$
(the lowest-numbered iteration at which any eigenstate is discarded) to $N$
(the highest-numbered iteration, and one at which we formally classify every
eigenstate as discarded). Here, $|l,M\rangle$ is one of the many-body
eigenstates discarded after iteration $M$ and $|e,M\rangle$, called
an ``environmental state,'' is any one of $d^{N-M}$ different possible simple
products of basis states for Wilson chain sites $M$ through $N-1$. For any $M<N$,
$|l,e,M\rangle$ is not an eigenstate of $H_N$ but the full density-matrix NRG
relies on a key approximation that $H_N |l,e,M\rangle \simeq H_M|l,M\rangle =
E_{l,M}|l,M\rangle$ (where all energies are measured relative to the ground
state of $H_N$).

Within the complete basis, the thermal equilibrium density matrix for the entire
system composed of the impurity and $N$ Wilson chain sites is diagonal and can
be written (after tracing out the environmental states)
\begin{align}
\label{eq:density_matrix}
\rho = \left( \begin{array}{ccccc}
\rho_{M_0} & 0 & & 0 & 0 \\
0 & \rho_{M_0 + 1} & & 0 & 0 \\
  & & \ddots &  & \\
0 & 0 & & \rho_{N-1} & 0 \\
0 & 0 & & 0 & \rho_N \\
\end{array} \right) ,
\end{align}
where $\rho_M$ is an $n_M^{\disc}\times n_M^{\disc}$ diagonal matrix having
matrix elements
\begin{equation}
(\rho_M)_{ll'}
= \sum_{e,e'} \langle l, e, M \mid \hat{\rho} \mid l', e', M \rangle
= \sum_{e,e'} \delta_{l,l'} \delta_{e,e'} e^{-\beta E_{l,M}} / Z
= \delta_{l,l'} d^{N-M} e^{-\beta E_{l,M}} / Z,
\end{equation}
with $\hat{\rho}=Z^{-1} \exp(-\beta H_N)$,
$Z = \mathrm{Tr}\, \exp(-\beta H_N)
   = \sum_{M=M_0}^N \sum_l d^{N-M} e^{-\beta E_{l,M}}$,
and $\beta=1/k_B T$.

We seek to calculate the von Neumann entanglement entropy $S_e$ with respect
to the partition of the system into a subsystem $A$ consisting of the impurity
and Wilson chain sites $n = 0$, $1$, $\ldots$, $L-1$ and a subsystem $B$ made
up of the remainder of the Wilson chain; see Fig.\ 1(b) of the main text.
Tracing out the degrees of freedom in subsystem $B$ yields the reduced density
matrix
\begin{align}
\rho_A = \mathrm{Tr}_B (\rho) = \left( \begin{array}{ccccc}
\rho_{M_0} & 0 & & 0 & 0 \\
0 & \rho_{M_0 + 1} & & 0 & 0 \\
  & & \ddots & & \\
0 & 0 & & \rho_L & 0 \\
0 & 0 & & 0 & R^{\text{red}}_L \\
\end{array} \right),
\end{align}
where $R^{\text{red}}_M$ is the \emph{partial} reduced density matrix with
elements $R^{\text{red}}_M(k,k')$ indexed by states $k$ and $k'$ \emph{kept}
(not discarded) after iteration $M$. $R^{\text{red}}_M$ can be
obtained from $R^{\text{red}}_{M+1}$ via reverse iteration along the Wilson chain
starting at $M=N-1$, as detailed in Eq.\ (30) of Ref.\ \onlinecite{Nghiem:14}.
Diagonalization of $R^{\text{red}}_L$ yields $n_L^{\kept}$
eigenvalues of $\rho_A$ that can be combined with the values
$(\rho_M)_{ll}$ for $M_0 \le M \le L$ to construct the full set of eigenvalues
$\{ \lambda_a \}$. Finally, one can compute the
entanglement entropy 
\begin{equation}
S_e = - \text{Tr}_A(\rho_A \ln \rho_A) = - \sum_a \lambda_a \ln \lambda_a .
\end{equation}

\subsection{Extraction of a characteristic temperature scale $T^*$}
\label{subsec:T*}

Figure 4 of the main paper shows that for $r>0$, the entanglement entropy
in each phase (Kondo and local-moment) scales as a function of $R/R^*$,
where $R^* = 1/(k_F T^*)$. For the purposes of this figure, we have
extracted the characteristic scale $T^*$ for any $J\ne J_c$ from the
temperature dependence of $\chi_{\imp}(T)$, the impurity contribution to
the uniform magnetic susceptibility. We define $4T^*$ to be the
temperature at which $T\chi_{\imp}$ reaches the midpoint between its
critical value (the one that persists to $T=0$ at $J = J_c$)
and its zero-temperature limit \cite{Gonzalez-Buxton:98-2} of $1/4$
(for $J<J_c$) or $r/8$ (for $J>J_c$). This temperature is taken to be
$4T^*$ (rather than $T^*$, say) so that for $r\to 0^+$ where
$4T^*\chi_{\imp}(4T^*)\to 0.125$, $T^*$ smoothly approaches the metallic
($r=0$) Kondo temperature, normally given the empirical definition
$T_K\chi_{\imp}(T_K) = 0.0701$ \cite{Krishna-murthy:80-2}.

\begin{figure}
\centering
\includegraphics[width=0.6\linewidth]{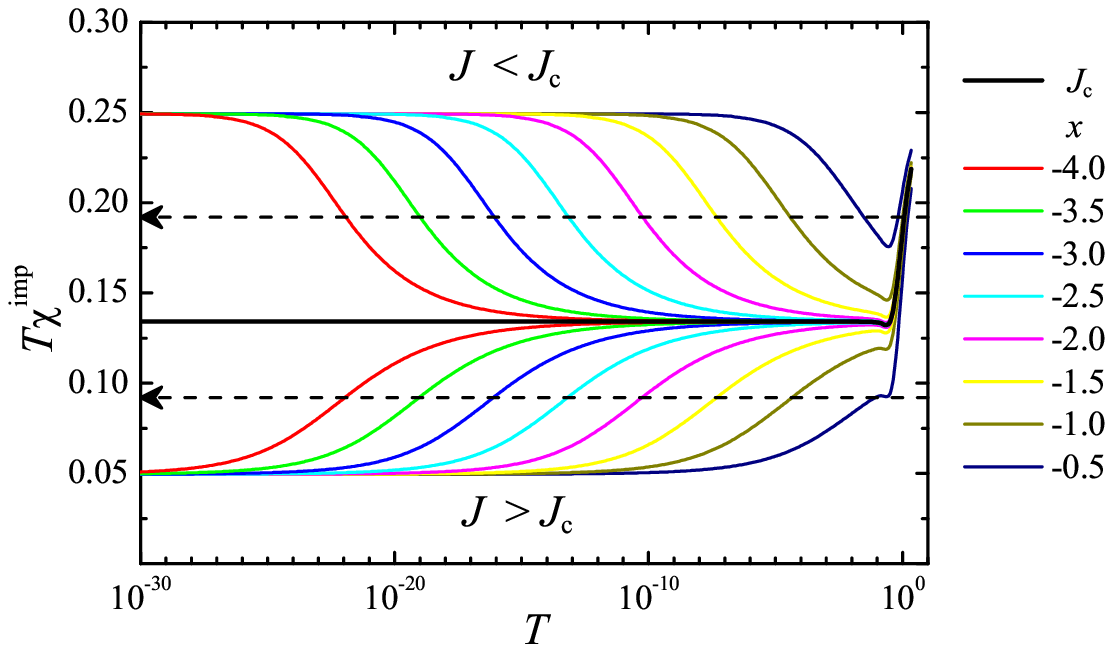}
\caption{\label{fig:T*}%
Extraction of the characteristic temperature scale $T^*$ from
$\chi_{\imp}(T)$, the impurity contribution to the uniform magnetic
susceptibility. For each value of $J$, $4T^*$ is defined to be the
temperature at which $T\chi_{\imp}(T)$ (solid curves thin solid lines)
reaches the midpoint (horizontal dashed line) between its $T\to 0$
limiting value for that $J$ (namely, $1/4$ for $J<J_c$, $r/8$ for
$J>J_c$) and the corresponding limiting value for $J=J_c$ (solid line).
Data shown are for band exponent $r=0.4$ and $J=(1\pm 10^x)J_c$ with
the values of $x$ shown in the legend.
}
\end{figure}

\subsection{Entanglement entropy as a function of Kondo coupling $J$}
\label{subsec:J-variation}

\begin{figure}
\centering
\includegraphics[width=0.6\linewidth]{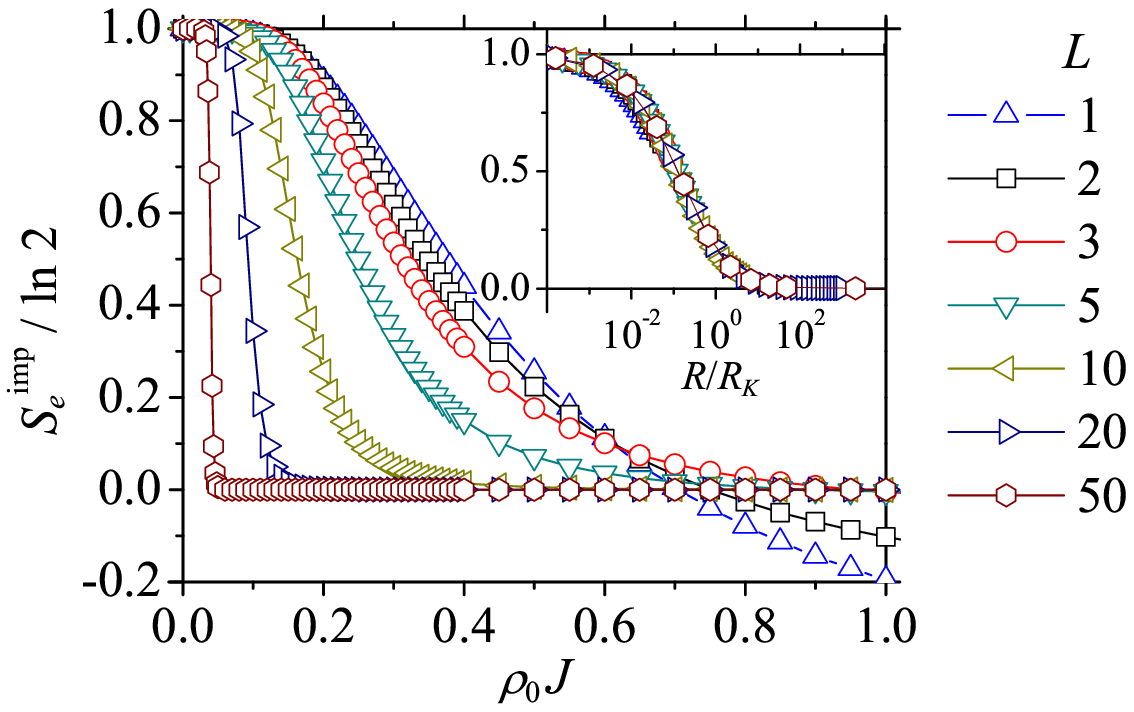}
\caption{\label{fig:SevsJ}%
Impurity entanglement entropy $\Sei$ vs dimensionless Kondo coupling
$\rho_0 J$ for band exponent $r=0$, discretization parameter
$\Lambda = 3$ and different partition sizes $L$.
The inset shows the collapse of curves for different $L$ values when the
data spanning $\rho_0 J\le 0.3$ are replotted as $\Sei$ vs $R/R_K$.
}
\end{figure}

The main paper presents results for $\Sei(J,L)$, the smoothed
(three-point-averaged) impurity contribution to the entanglement entropy
as a function of the Wilson chain partition size $L$ for different fixed
Kondo couplings $\rho_0 J$. Figure \ref{fig:SevsJ} instead plots
$\Sei$ vs $\rho_0 J$ for the metallic case $r=0$ with each data set
representing a different fixed partition size $L$.
With increasing $J$, each partition shows a monotonic decrease of $\Sei$.
For very weak Kondo couplings $\rho_0 J\ll 1$, the impurity spin is
collectively screened by essentially the entire Wilson chain. The amount
of screening that takes place within the first $L$ sites of the Wilson
becomes ever smaller as $J\to 0^+$, so the impurity's entanglement with
chain sites $n\ge L$ approaches the full value $\ln 2$ for a spin singlet.

For the opposite limit $\rho_0 J \gg 1$, in the ground state of $H_N$ given
by Eqs.\ \eqref{H_M} and \eqref{H_M^chain}, the impurity is essentially
locked into a spin singlet with the on-site combination of conduction
electrons annihilated by the $f_{0\sigma}$ operator; chain sites
$1$, $2$, $\ldots$ $N-1$ behave like a free Wilson chain partitioned into
segments of length $L-1$ and $N-L$. As a result, the impurity contribution
to the entanglement entropy can be written $\Sei(J,L,N) =
S_e(J,L,N) - S_e^{(0)}(L,N) \simeq S_e^{(0)}(L-1,N-1) - S_e^{(0)}(L,N)$,
where $S_e^{(0)}$ is the entanglement entropy of a chain of length $N$
partitioned into $L$ and $N-L$ sites.
After making $N$ very large and performing a three-point average, the
smoothed impurity entanglement entropy $\Sei(J,L)$ defined in
the main paper is negative for $L\lesssim L_{\Lambda}$---over which range
$\Sea(L,N\gg L/2)$ grows with increasing $L$---and rapidly approaches
zero for $L\gtrsim L_{\Lambda}$. For the value $\Lambda = 3$ used to
produce Fig.\ \ref{fig:SevsJ}, $L_{\Lambda} \simeq 1$ and negative
$\Sei$ values are found only for $L\lesssim 3$.

As $L$ is increased, the crossover in $\Sei$ from $\ln 2$ toward zero
takes place more sharply and centered around a smaller value of $\rho_0 J$.
This is another manifestation of the notion presented in the main text
that $\Sei$ drops once the radius $R$ of subsystem $A$ exceeds the
characteristic size $R_K$ of the Kondo screening cloud. The inset of Fig.\
\ref{fig:SevsJ} replots the data for $\rho_0 J \le 0.3$ as a function of
$R/R_K$, where each $L$ curve corresponds to fixed value of
$R = c \Lambda^{L/2}/k_F$ (with $k_F$ being the Fermi wave vectors and
$c$ a constant of order unity) and points within a curve arise from
a decrease with increasing $J$ of $R_K \sim 1/(k_F T_K)$.
Whereas in the main paper, the Kondo temperature $T_K$ was deduced from
the impurity contribution to the magnetic susceptibility via the
conventional definition $T\chi_{\imp}(T_K) = 0.0701$ \cite{Krishna-murthy:80-2},
in Fig.\ \ref{fig:SevsJ} we instead employed the perturbative definition
 \cite{Hewson:93-2}
\begin{align}
\label{T_K}
k_B T_K \sim D \sqrt{\rho_0 J}\exp[-1/(\rho_0 J) + O(\rho_0 J)].
\end{align}
The collapse of all curves except those for $L = 1$ and $2$ (which are
anomalous for reasons discussed in the preceding paragraph) is
consistent with the existence of a universal scaling function
$\Sei(J,R) = f_0(R/R_K)$, as also argued on the basis of the data
presented in the main paper.

Similar behavior can be seen in plots (not shown) of $\Sei$ vs $\rho_0 J$
at fixed $L$ for pseudogapped hosts (i.e., $r>0$). The data in each phase
(Kondo or local-moment) can be collapsed by plotting $\Sei$ against $R/R^*$,
where the crossover length scale $R^*= 1/k_F T^*$. While $T^*$ can be
determined from $\chi_{\imp}(T)$ via the operational procedure laid out in
Sec.\ \ref{subsec:T*}, for values of $J$ sufficiently close to $J_c$, good
collapse can be achieved by instead using the asymptotic expression
\begin{align}
\label{eqn:Tstar}
T^* \propto |J-J_c|^{\nu},
\end{align}
where the numerical value of the correlation length exponent $\nu$ has
a nontrivial dependence on the band exponent $r$ \cite{Ingersent:02-2}.

\subsection{Fixed-point entanglement entropy vs $r$}
\label{subsec:Se_fixed-point}

This section provides more details of the $r$ dependence of the impurity
entanglement entropy at each renormalization-group fixed point, as well as the
manner in which $\Sei$ approaches its value at each of the stable fixed points.

The results in the main paper show that, whereas $\Sei=0$ at the weak-coupling
fixed point, the impurity entanglement entropy takes nontrivial, $r$-dependent
values at the Kondo-destruction quantum critical point and at the Kondo fixed
point. The fixed-point values of $\Sei$ can be obtained from many-body NRG
calculations by taking the limit $R\ll R^*$ (for the unstable critical point) or
$R \gg R^*$ (for the stable Kondo and local-moment fixed points). The Kondo
fixed-point value of $\Sei$ can also be calculated using the single-particle
method outlined in Sec.\ \ref{subsec:WC_calculation} as the difference of
$\Sea$ for a free Wilson chain with and without the first site frozen due
to the formation of a local spin singlet with the magnetic impurity. That
the many-body and single-particle approaches yield numerical values in
excellent agreement provides a valuable check on the accuracy of the full NRG
results.

\begin{figure}
\centering
\includegraphics[width=1.0\linewidth]{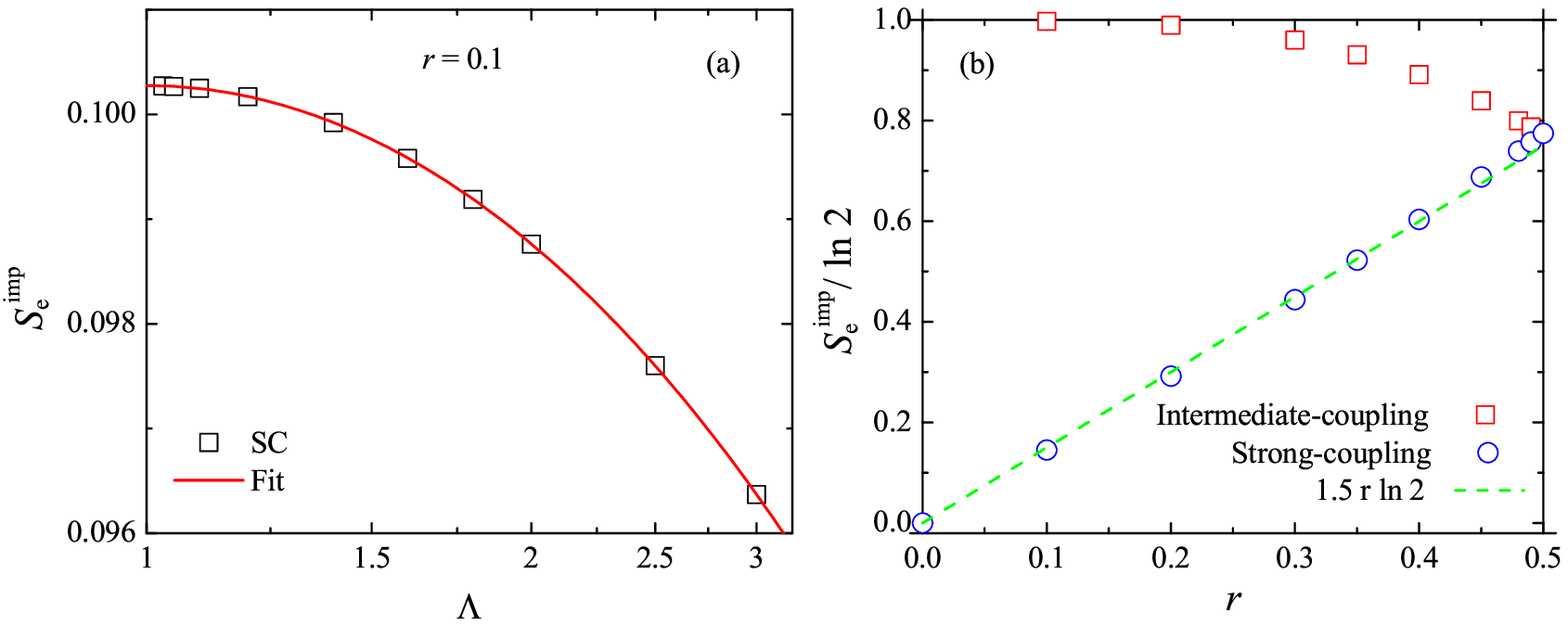}
\caption{\label{fig:Se-fixed-point}%
Impurity entanglement at the quantum critical point and the Kondo fixed
point:
(a) Kondo fixed-point value of $\Sei$ (symbols) vs $\Lambda$ (on a log scale)
for band exponent $r = 0.1$. A polynomial fit (solid line) is used to
extrapolate $\Sei$ to the continuum limit $\Lambda = 1$.
(b) Extrapolated $\Lambda=1$ values of $\Sei$ at the critical point (squares)
and at the Kondo fixed point (circles) vs band exponent $r$, along with a
heuristic fit $\Sei = \frac{3}{2} r \ln 2$ (dashed line).
}
\end{figure}

In order to remove discretization effects, fixed-point values of $\Sei$
were calculated for values of $\Lambda$ between $1.01$ and $3$, then
fitted with a polynomial function of $\ln\Lambda$, allowing extrapolation
of $\Sei$ to the continuum limit $\Lambda =1$, as illustrated in
Fig.\ \ref{fig:Se-fixed-point}(a).

\begin{table}[h]
\begin{tabular}{lll}
\multicolumn{1}{l}{$\ r$}
  & \multicolumn{1}{c}{$\alpha$(LM)}
	  & \multicolumn{1}{c}{$\alpha$(K)} \\ \hline 
0.0  &  0        &  1.000(3)  \\
0.2  &  0.38(3)  &  0.800(5) \\
0.25 &           &  0.750(4) \\
0.3  &  0.58(5)  &  0.693(8) \\
0.33 &           &  0.630(7) \\
0.4  &  0.79(4)  &  0.399(5) \\
0.45 &  0.89(5)  &  0.199(2) \\
0.5	 &  1        &  0
\end{tabular}
\caption{\label{tab:alpha}%
Values of the exponent $\alpha$ defined in Eq.\ \eqref{eq:alpha}
for different band exponents $r$, as determined in the local-moment (LM)
and Kondo (K) phases. A number in parentheses denotes the estimated
nonsystematic error in the last digit. Values without error estimates are
assumed rather than computed.}
\end{table}

\begin{figure}
\centering
\includegraphics[width=0.45\linewidth]{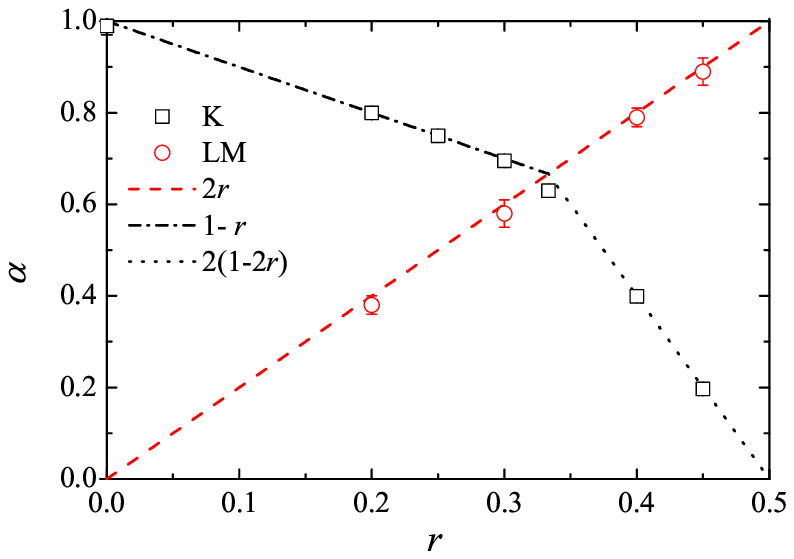}
\caption{\label{fig:alphas}%
Values of the exponent $\alpha$ defined in Eq.\ \eqref{eq:alpha}
for different band exponents $r$, as determined in the local-moment (LM,
circles) and Kondo (K, squares) phases, along with lines showing the
functions $\alpha=2r$, $1-r$, and $2(1-2r)$.
}
\end{figure}

Extrapolated values of $\Sei$ are shown in Fig.\ \ref{fig:Se-fixed-point}(b).
As $r$ increases from $0$, the critical value of $\Sei$ decreases from
$\ln 2$ while the Kondo fixed-point value increases almost linearly from $0$.
The two fixed-point values meet at $r=\half$, the band exponent
at which the quantum critical point merges with the Kondo fixed point.
(No quantum critical point exists for $r>\half$
\cite{Gonzalez-Buxton:98-2}.)
A weak superlinear variation can be seen when the Kondo entanglement
entropy is compared with a heuristic fit $\Sei = \frac{3}{2} r \ln 2$
[dashed line in Fig.\ \ref{fig:Se-fixed-point}(b)].
This superlinear behavior is somewhat unexpected since thermodynamic
properties at strong coupling have been shown to exhibit a strictly
linear variation with $r$ \cite{Gonzalez-Buxton:98-2}.

Insets in Fig.\ 4 of the main paper demonstrate that $\Sei$ has a
power-law-decaying tail in both the local-moment and Kondo phases, namely,
\begin{equation}
\Sei(J,R)-\Sei(J,\infty) \propto (R/R^*)^{-\alpha}
  \quad \text{for } R\gg R^*.
\label{eq:alpha}
\end{equation}
Fitted values of $\alpha$ are listed in Table \ref{tab:alpha} and plotted
in Fig.\ \ref{fig:alphas}. To within the estimated numerical uncertainty,
the extracted exponents are consistent with $\alpha = 2r$ for $J<J_c$ and
$\alpha = \text{min}(1-r, 2-4r)$ for $J>J_c$. These expressions coincide
with twice the exponent of the leading irrelevant operator at the
local-moment and Kondo fixed points, respectively; see Eqs. (4.7) and
(4.10) in Ref.\ \onlinecite{Gonzalez-Buxton:98}. This is consistent with
the natural interpretation that the power-law tails are associated with
the renormalization-group flow toward the stable fixed point in either phase,
leading to the expectation that the exponent $\alpha$ is a characteristic
of that fixed point. 

It is probable that the departure of $\Sei$ from its value on the
critical plateau is also described by a power-law behavior, i.e.,
\begin{equation}
\Sei(J,R)-\Sei(J_c,\infty) \propto (R/R^*)^{\alpha'}
  \quad \text{for } R\ll R^*,
\label{eq:alpha'}
\end{equation}
where one would expect $\alpha'$ to be positive and a characteristic property
of the Kondo destruction critical point (and, hence, likely to have a
nontrivial $r$ dependence).
However, numerical uncertainty in the value of the critical value
$\Sei(J_c,\infty)$ impedes reliable determination of $\alpha'$.


\begin{thebibliography}{1}
\bibitem{L=0}
  Reference \onlinecite{Pixley:15} studied $\Sei(J,0,N)$, which is not directly
  comparable with the extrapolation to $L=0$ of the smoothed quantity
  $\Sei(J,L)$ discussed in this work.
\end{thebibliography}
\end{document}